\documentclass[10pt,preprint]{aastex}
\usepackage{color}
\begin{document}

\title{An Alternative Explanation of the Varying Boron to Carbon in Galactic Cosmic rays}
\author{David Eichler$^{*}$}
\altaffiltext{*}{Dept. of Physics, Ben-Gurion University, Be'er-Sheba 84105, Israel}

\begin{abstract}
It is suggested that the decline with energy of the boron to carbon abundance ratio in Galactic cosmic rays is due, in part, to a correlation between the maximum energy attainable by shock acceleration in a given region of the Galactic disk and the grammage traversed before escape.  In this case the energy dependence of the escape rate from the Galaxy may be less than previously thought and the spectrum of antiprotons becomes easier to understand.  

\end{abstract}


Acceleration by strong shocks (compression ratio of 4 or more) is expected theoretically to produce $E^{-2}$ spectra during most of the life of a supernova remnant (Ellison and Eichler, 1985), yet the Galactic cosmic ray spectral index is $E^{-2.7}$.  Some of the discrepancy  may be attributed to  the fact that the last stages of shock acceleration  are  with shocks that have compression ratio less than 4. Mostly however,  the discrepancy is usually attributed to an energy dependent escape ratio, as is evidenced by  the secondary to primary ratio, which decreases with energy. But the recent announcement of an antiproton spectrum that is identical to  primary spectrum raises questions about this, because antiprotons\footnote{unless the result of dark matter decay or annihilation} are also secondaries. Moreover, the positrons, which may also be secondaries, also have an identical spectrum to the primaries.  Cowsik and    Madziwa-Nussinov  (2016) have recently  suggested that the escape rate from the Galaxy is in fact energy independent, and that the primary source spectral index is really 2.7 and not 2. However, it would then be unclear why the  boron to carbon ratio  is energy dependent. To address this question, Cowsik and coworkers have suggested that most of the boron is made in  ``nested'' leaky boxes that encompass the production sites of cosmic rays and that the escape rate from the nested leaky box, though not from the Galaxy, has the required energy dependence, but it is unclear why the one but not the other would be energy dependent. 

In this paper, I suggest an alternative reason for the decrease of the B/C ratio with energy, which does not demand an energy dependent escape rate. While the idea is somewhat speculative, I believe it should not be overlooked, even if only partly true, because it may play a role in determining the values of measured quantities. 

The Galactic disk is very thin. Most of the baryonic matter is concentrated within layers that  are of order 100 pc from the equatorial plane, and this is only about $10^{-2}$ of the Galactic radius R.  The limits of CR anisotropy at $E\gtrsim$ 1 TeV suggest that their streaming velocity is of order $10^{-4}$c or less, and this means that CRs do not stray far from their sources before escaping the disk (Parker, 1969).
{ See also D'Angelo et. al (2016) for an analysis of  diffusion in self-generated turbulence that is consistent with this conclusion.} Parizot and Drury (2000) have argued that cosmic ray secondary production  near the production sites of the primary elements can explain the evolution of Li, Be, and B in the Galaxy. The disk is also very inhomogeneous and different regions could have different densities, ionization fractions, different rates of star formation, different levels of Alfven wave turbulence, etc. The correlation lengths of these varying quantities could be smaller or larger than the disk thickness.  Accordingly, the cosmic ray escape rate from the disk, and the grammage of interstellar matter that they traverse before doing so, could differ among  different parts of the disk.  

The maximum energy to which a supernova shock can accelerate cosmic rays can also depend strongly on the Galactic environment in which the shock exists. Ion-neutral damping, for example, can severely limit $E_{max}$ (Bell, 1978). That cosmic rays appear to be made in regions in which there are grains suggests that warm regions of the interstellar medium are well represented in overall cosmic ray production.  Gamma rays from supernovae remnants display a wide variety of spectral indexes and cutoffs, and this illustrates the diversity of cosmic ray spectra that could be expected.

In this letter it is suggested that $E_{max}$ at the site of cosmic ray production is anti-correlated with the grammage traversed by a cosmic ray before it escapes the Galaxy (or the subregion thereof where most of the grammage is traversed). This would create an energy dependence in the B/C ratio even if the escape rate from the Galaxy is energy independent. The suggestion is quite general, but a specific version of it will be suggested following a more general description.  It is not claimed here that the escape rate  of cosmic rays from the Galaxy is entirely energy independent,  merely that that an energy-dependent escape rate need not be the only reason for an energy dependence in the B/C ratio.

Let us express the primary  CR production spectrum at a given site of production as $N_p(E,{\bf x})$ below $E_{max}$, where $N_p(E)=E^{-2-p} \, p \ge 0 $, and where $N_p(E,\bf x)$ vanishes at $E \ge E_{max}$.  The maximum energy $ E_{max}$ specific to that source is determined by any one of several physical considerations to be discussed later. The overall spectrum $ N_{p,T}(E, {\bf x})$ at any point {\bf x} in the Galaxy is then $N_p S(E_{max}\ge E)$, where $S(E_{max}\ge E)$ is the total source contribution in sources that allow acceleration up to or beyond  energy E. This fact can be written as
 
\begin{equation}
N_{p,T}(E,{\bf x}) = E^{-2-p}\int_E^{\infty} [dS(E_{max},{\bf x})/dE_{max}]dE_{max},
\end{equation}
  where $dS(E_{max},{\bf x})/dE_{max}$ is the relative strength of sources with a maximum energy  between $E_{max}$  and $E_{max}+ dE_{max}$ contributing to point {\bf x}. Note that the ``relative strength'' $S(E_{max},{\bf x})$ depends both on the number and strength of regions with maximum energy $E_{max} $ as well as their distance from point {\bf x}, but what is relevant here is just the net  source strength.
 For simplicity assume that $[dS(E_{max},{\bf x})/dE_{max}]$  is a power law $E_{max}^{a-1}dE_{max} $, [i.e. $S(E\ge E_{max})  \propto E_{max}^{a}$] up a ``maximum'' $E_{max}$ $\equiv E^*$, and vanishes at $E_{max}> E^*$. Ignore, also for simplicity, possible variation of $a$ with location {\bf x}; i.e.
 
 \begin{eqnarray}
 dS(E_{max},{\bf x}) \propto  E_{max}^{a-1}dE_{max};  \,&E_{max}\le E^*& \nonumber\\
         \propto 0; \,\,\,\,\,\,\,\,\,\, \,\,\,\,\,\,\,\,\,\,\,\,\,\,\,\,\,\,\,&\, E_{max}> E^* &
 \end{eqnarray}
 
 then
 
 \begin{equation}
 N_{p.T}(E,x)\propto E^{-2-p}[{E^*}^a - E^a] , \,\,\,  E\le E^*
 \end{equation}
 
 For secondaries, such as boron, the overall spectrum $N_{s,T}$ is the sum of secondaries produced by primaries that were produced with the various values of $E_{max}$. The contribution from each $E_{max}$ is proportional to the average grammage $G(E,E_{max})$ traversed by primaries of energy $E$ that are produced with maximum energy $E_{max}$.  Let us further assume that the relative strength $S(E_{max})$ of various points of the Galaxy contributing to $N_T$ is the same for secondaries as for primaries, then the total secondary spectrum $N_{s,T}$ is given by 
 
 \begin{equation}
N_{s,T}(E) \propto  \int_E^{\infty}\left[\int_{E}^{max}N(E')K(E',E)G(E',E_{max})[dS(E_{max})/dE_{max}] dE' \right] dE_{max}
\label{nsecondary}
   \end{equation}
Here $K(E,E')$ is the multiplicity of secondaries at energy $E$ produced by a collision with primary energy $E'$. For a spallation product, if we take the ``energy'' E to mean energy per nucleon, $K(E,E')$ is to a good approximation $\delta(E'-E)$ so equation (\ref{nsecondary}) reduces to

\begin{equation}
N_{s,T}\propto \int_E^{\infty}N(E)G(E, E_{max})[dS(E_{max})/dE_{max}] dE_{max}
\end{equation}
To simplify still further, let us now write $G$ as a function only  of $E_{max}$ but not of E, which allows for the possibility that $G$ is correlated with $E_{max}$, while not depending on energy $E$. Suppose, for example, that 
$G( E_{max})\propto E_{max}^{-\beta}$. Then primary spectra such as carbon would be as before 
 while boron would be
 \begin{equation}
 N_{s.T}(E,x)\propto E^{-2-p}[{E^*}^{a-\beta }- E^{a-\beta}] /[a-\beta]
 \end{equation}
 Now if $a > 0$ and $a-\beta <0$, the expression $E^{-2-p}[{E^*}^a - E^a]  $ is dominated by the first term in the brackets,  because $E^*>E$, while $ E^{-2-p}[{E^*}^{a-\beta }- E^{a-\beta}]  $ is dominated by the second term.  If $E^* \gg E$. then, to  a good approximation,
  \begin{equation}
 N_{p.T}(E,x)\propto E^{-2-p}
 \end{equation}
 while
  \begin{equation}
 N_{s.T}(E,x)\propto E^{-2-p}[ E^{a-\beta}].
 \end{equation}
  That is to say, the primaries can have the same spectrum as the production spectrum, while the secondaries have a steeper one, even though the escape rate is energy independent.  This happens when most of the CRs are made in regions where $E_{max}$ is nearly $E^*$, while most of the secondaries are produced in a minority of regions  that just happen to have low $E_{max}$ and a high  value of  grammage $G$. 
  
   The   case could be made that cool and warm (as opposed to hot) regions, where most of the refractory elements would be locked up in grains and where the density would be highest, contribute a small fraction of primaries and a large fraction of low energy secondaries.  Refractory elements have a higher relative abundance in the cosmic rays than volatile ones (Ellison, Drury, and Meyer, 1997). This is attributed to the fact that the grains, being charged but very massive  relative to protons, therefore have a higher rigidity than protons, which guarantees them entry into the Fermi acceleration process. However, in the warm phase, most heavy elements are believed to be locked up in grains. If most of  the heavy elements are therefore injected into the diffusive shock acceleration  (DSA) process, the refractory elements would be enhanced by far more than the observed factor of  $\sim 4$.  
   Thus, it could be argued that  most of the cosmic rays do not come from the cool or warm phase, whereas most of the refractories do. (The argument is not airtight, because it could be that the grains are so massive  that, even if charged, they are not turned around by the shock, but this would require fine tuning.)
   
   One can apply similar reasoning to secondaries: If CRs are blown out of the disk in collective outflows from multiple supernovae, one might expect that the grammage $G$ they traverse is proportional to the density but inversely proportional to  the convection velocity u, while u might anticorrelate as density n as well. Thus $G$ may depend on density n to a power that exceeds unity - i.e. faster than a linear dependence. This could mean (see below, after further explanation) that most of the secondaries,  but only a small fraction of the primaries,  are made in the warm phase.
   
Now consider the quantity $E_{max}$. In the following discussion, a specific example is given of how $E_{max}$ might anti-correlate with  $G$.  $E_{max}$ is limited (e.g. Bell 1978; Drury, Duffy and Kirk, 1996 - hereafter DDK) by the condition that ion-neutral damping not eliminate the  waves necessary to confine the accelerated CR near the shock front. I now summarize the calculations of DDK.  Suppose that  the wave energy density is dissipated by ion neutral damping at the rate of $ \tau_d^{-1} \delta B^2 /4\pi $, where $\tau_d$ is the wave damping time. When this exceeds the rate of energy gain the waves cannot survive. The rate of wave energy gain per unit volume in the frame of the fluid, $\dot U_W$, due to the force per unit volume exerted by the CR on the incoming fluid, $\cal F$, is given by $v_{ph}\cal F$, where $v_{ph}$ is the phase velocity of the waves and  ${ \cal F }= dP_{CR}/dx$.

Now at a given energy E, define a quantity $\phi$  with units of energy flux as 
\begin{equation}
\phi = D(E)dP_{CR}/dx
\end{equation}
 where D(E) is the diffusion coefficient of energy E. 
D(E)  is given, in turn, by
\begin{equation}
D(E) \simeq \eta  [B^2/\delta B^2] r_g c/3 ,
\end{equation}
where $r_g$ is the CR gyroradius, and $\eta \sim 1$ and is henceforth dropped.  When there is steady state with no escape upstream, $\phi = uP_{CR}$ upstream of the shock.

So the condition that the wave growth be positive, $[\delta B^2/4\pi] \tau_d^{-1}\le v_{ph}dP_{CR}/dx$, can be written as 
\begin{equation}
 \tau_d^{-1}  \le   v_{ph}\phi \left[{4\pi \over B^2}\right][3eB/Ec]
\end{equation}

In order for shock acceleration to work, the diffusive   flux $\phi$ of particles escaping upstream must not exceed  the convective flux $uP_{CR}$ in the shock frame by more than the downstream convective losses $u_+ P_{CR,+}$, where $u_+ \equiv u_s/r$ is the downstream velocity in the shock frame,  $r$ is the compression ratio, $P_{CR,+}$ is the post shock CR pressure, and, since $P_{CR}$  at the upstream free escape boundary is probably much less than  $(u_+/u)P_{CR,+}$, we can just say $\phi \le  u_+ P_{CR,+}$.
\begin{equation}
\phi \le  u_+ P_{CR,+}
\end{equation}
whence
\begin{equation}
 \tau_d^{-1}  \le 3 v_{ph}u_+ P_{CR,+}\left[{4\pi \over B^2}\right]/r_g c 
 \label{condition}
\end{equation} 

 The damping rate is (DDK)
 \begin{equation}
 \tau_d^{-1} = \left({\omega^2\over \omega^2 + (n_i /n\tau_c)^2}\right) \tau_c^{-1}\le \tau_c^{-1}.
 \label{taud}
 \end{equation}
 where 
 \begin{equation}
 \tau_c^{-1} = [8.4 \cdot 10^{-9}][{T\over 10^4 K}]^{0.4} n \,\rm cm^3 s^{-1}
 \label{tauc}
 \end{equation}
   (e.g. Parker, 1969; Kulsrud and Cesarsky,  1971) is the frequency with which a given ion collides with a  neutral, and where $n$ is the number density of the neutrals. The time over which the neutrals are dragged along with the ions is $({n\tau_c\over n_i}) $.

      Assuming the shock acceleration to be efficient, with an $E^{-2}$ differential spectrum,  $P_{CR} \sim \rho u_s^2/ln({E_{max}\over E_{min}})$, we we use  $P_{CR} \equiv \rho u_s^2/ln\lambda$ for brevity, and rewrite equation (\ref{condition}) as 
\begin{equation}
 \tau_d^{-1}  \le  3 v_{ph}u_+[{4\pi \rho\over B^2}] u_s^2/ r_g c ln\lambda = ({3\over r})v_{ph} u_s^3/r_g c v_A^2 ln\lambda.
 \label{recondition}
\end{equation}
Note that if the spectrum is steeper than $E^{-2}$,  $ln\lambda$ is effectively raised, because the CR  pressure at $E_{max}$ is lowered below $\rho u_s^2/ ln({E_{max}\over E_{min}})$.

Assuming  the waves are gyroresonant with the CR that generate them, then $v_{ph} =r_g \omega$ where $\omega = kv_{ph}$ is the wave frequency, and, with equation (\ref{taud}),  equation  (\ref{recondition}) can be written    

 \begin{equation}
({\omega^2\over \omega^2 + (n_i /n\tau_c)^2}) \tau_c^{-1} \le( {3\over r}) \omega u_s^3/c v_A^2 ln\lambda.
 \label{condition2}
\end{equation}

An upper limit on particle energy, or equivalently, on $r_g$, is obtained only if 

\begin{equation}
\omega^2 \gtrsim (n_i /n\tau_c)^2 ,
\label{omegagtr}
\end{equation}
in which case  the condition reduces to 
   \begin{equation}
r_g    = E/qB \le  v_{ph}\tau_c ( {3/r})  u_s^3/c v_A^2 ln\lambda  \equiv  E_{max}/qB .
\label{rg}
\end{equation}
(If  $\omega^2 \ll (n_i /n\tau_c)^2$ then condition (\ref{condition2}) becomes a lower limit on E, which is meaningless if particles are unable to reach this lower limit,  and in any case empty if they are able to.)

 Equation   (\ref{rg}) sets an upper limit, $E_{max}$, to the energy of
 \begin{equation}
 E_{max} \sim  (v_{ph}/v_A)(u_s/c)^3  (T/10^4 K)^{-0.4}
(n/1 cm^{-3})^{-1}
(n_i/1 cm^{-3})^{1/2}
(10/ln \Lambda) 10^{19} eV,
\label{emax2}
\end{equation}
which for $u_s = 10^3 km/s$ is of order 300 GeV.

 So, for example, if CRs are convected out of the Galactic disk at a velocity of  $10^{-4}\beta_{-4} c$, independent of energy (Parker 1969),  then they would traverse the disk thickness, $\sim 100$ pc, in about $3 \beta_{-4}  ^{-1}$ Myr, and would traverse $5 \beta_{-4}{(n/1 \, \rm cm^{-3})}$ g cm$^{-2}$. Considering that  $(v_{ph}/30 \, \rm km s^{-1})(B/3 \mu G)(T/10^4 K)^{-0.4}  (10/ln \lambda)$ would be only somewhat less than if not greater than unity, as its individual  factors are of order unity, their spectrum would extend up to at least several hundred $(n/1 \, \rm cm^{-3})^{-1}$ GeV or so. So the value of $\beta$ as defined before equation (4) is -1. 

Now condition (\ref{omegagtr}) itself sets an upper limit on E of 

\begin{equation}
E/qB \le v_{ph}\tau_c [{n\over n_i}] = v_{ph}/<\sigma v>n_i
\label{eoverqB}
\end{equation}
giving
\begin{eqnarray}
E_{max*}
 &=& 70 [B/3 \mu G]^{2}[T/10^4 K]^{-0.4}[n_i/1 cm^{-3}]^{-3/2}[v_{ph}/v_A] \,{\rm GeV}
\label{emax*}
\end{eqnarray}
While the phase velocity $v_{ph}$ is usually taken to be the Alfven velocity $v_A$,  $v_{ph}$, in the presence of a strong driving force, may  be much greater than $v_A$ (Fiorito, Eichler, and Ellison, 1990), and as high as $u_s/4 \gg v_A$ without shutting off shock acceleration. Note that $E_{max*}$ as defined in equation (\ref{emax*}) is not an upper limit to shock acceleration, but rather an upper limit to the energy range at which  shock acceleration would be limited by ion-neutral damping.

 To summarize, $E_{max}$ is either a) limited by ion-neutral damping to $E_{max} \le E_{max*}=70 [{v_{ph}\over v_A}][{B\over 3 \mu G}]^{2}[{T\over 10^4 K}]^{-0.4}[{n_i\over 1 cm}]^{-3/2}$  or b)  is not limited at all by ion neutral damping. It is in any case limited by size and age of the supernova remnant to about $10^4$ or $10^5$ GeV (Lagage and Cesarsky, 1983) which is well below the knee. In the hot phase of the ISM, assuming $T= 7 \cdot 10^5$ K and $n_i=3 \cdot 10^{-3}$, equation (\ref{emax*}) states that $E_{max}$ could attain values as high as $5 \cdot 10^4 v_{ph}/v_A$ GeV with ion-neutral damping allowing eventual escape before severe adiabatic losses set in.  
 
 That the spectral index of -2.7 remains nearly constant until well above the limit established by the diffusive shock acceleration   (DSA) limit claimed by Lagage and Cesarsky (1983) remains a puzzle, though it has been claimed (Jokipii, private communication) that shock drift can surpass the DSA limit. But this is in any case a puzzle for any supernova remnant, even if CR trapping is not limited by ion-neutral damping.  It may be that, for some still not understood reason,  the number of supernova with $E_{max}\gtrsim 10^4$ GeV [or the contribution from shock drift)] declines as $E_{max}^{a};\, a \sim -0.7$.
 
  Recall  that the total  grammage $G \propto (n+n_i)t$ traversed  by a cosmic ray before convected out of the disk in time $t$   goes as $n^{\alpha}$,  where, if a) the material is mostly neutral ($n+n_i $) is non-decreasing with n, and b) t is nondecreasing with n, then $\alpha\ge 1$.  
 Now suppose that  dense cloud regions have neutral densities $n\lesssim 10^{3}\rm cm^{-3}$.
  Equation (\ref{emax2}) then suggests that $G\propto E_{max}^{-\beta}$, where $\beta \ge 1$.  This implies that even if most of the contribution to the CR we observe is made at $E_{max} \gg $ 1 TeV, the possibility exists that a small fraction is made in dense regions where $n\gtrsim 10^2$ cm$^{-3}$ and $E_{max}$  ranges from several GeV to over 100 GeV. 
 
The question of adiabatic losses in incompletely ionized media is pertinent to the relative weights of various contributors to the Galactic CR pool: Although highly ionized media allow acceleration to higher energy,  they may, by the same token, require more adiabatic losses once the CR are accelerated.  Partially ionized media on the other hand, as $E_{max}$ decreases with the decreasing expansion velocity of the supernova remnant, release  CR at $E\ge E_{max}$ that may have been accelerated at earlier stages. This suggests that incompletely ionized regions of the ISM such as the warm ISM and dense regions of new star formation, may be favorably represented relative to highly ionized  regions, such as the hot ISM, simply because they more effectively release the CR produced within them with less adiabatic loss.

 Although the hot ISM  makes up most of the volume, most of the supernovae may occur in dense regions of newly forming stars, and this is another reason that incompletely ionized parts of the ISM may be favorably represented as CR sources relative to the hot ISM.  However such regions are likely to have a high concentration of young, UV emitting stars and  supernovae, and the interstellar gas in them is likely to make sudden transitions from neutrality to a state of high ionization, so they defy simple parametrization. With that in mind, consider that the density in the star forming region can be as high as $10^2 -10^3$ cm$^{-3}$.  Suppose a collective blast from multiple supernova forms an expanding superbubble.  The Stromgren sphere of photoionization from the young stars is overtaken by the forward shock wave of the expanding superbubble over a time scale of 0.1 Myr (Gupta et al, 2016), and after that, the shock expands into dense mostly neutral medium, where by  equation (\ref{emax2}) $E_{max}$ can be as low as several GeV, and where the grammage traversed by any CRs trapped at the shock can be as high as tens of grams per cm$^2$,  i.e. thick target. Because the grammage can be such a steep function of density, steeper than the dependence of $S(E_{max})$ on $E_{max}$, then it is possible that  these dense regions  spawn most of the secondaries while they contribute little to the primaries.
 
 The question of quantifying  the distribution $S(E_{max})$, i.e. how the parameters that decide $E_{max}$ are distributed in the Galactic disk, is beyond the scope of this paper, however, it will be the subject of future research. It is possible that interesting constraints come from the fact that refractory elements have the same slopes  as volatiles, which set a limit on how much the warm phase contributes to the Galactic cosmic rays.
 
  The question might arise why the Galactic CR spectral index is -2.7 rather than -2.  This is an old question and the usual suggested answers are that most of the volume swept out by supernova shocks is when the shocks are no longer at their strongest, and/or that the escape rate from the galaxy is rigidity-dependent. But an additional possibility is that the spectral index  $a$ of $S(E_{max})$, is negative. This would still beg the question of why the antiprotons have a flatter spectrum than boron. Cowsik and Madziwa-Nussinov (2016) suggest that, at primary energies necessary for antiproton production, the grammage traversed in the nested leaky box (where they hypothesize most of the boron production at lower energies takes place) is less than in the disk at large.  Perhaps the spirit of this suggestion can be adapted to the concept of correlation between $E_{max}$ and G in view of   equations (\ref{emax2}) and (\ref{emax*}): These equations suggest that, in the warm phase, beyond sufficiently high energy to make 100 GeV antiprotons, $\sim 1$ TeV, the exponent $a$, which characterizes   the dependence $S(E_{max})\propto E_{max}^a$,   itself changes as a function of $E_{max}$; i.e.  $S(E_{max})$ depends on $E_{max}$ at $E_{max} < E_{max*}$  and that, at higher $E_{max}> E_{max*}$, $S(E_{max})$ ceases to be $E_{max}$-dependent.\footnote{ until of course the shock-radius-limiting energy is reached.} If this were the case, and still assuming that escape from the Galaxy is via convection and therefore not significantly energy-dependent, the spectral index of antiprotons would be  the same as that of the primaries, as observed. This would predict some flattening of the secondary to primary ratio  near and above 1 TeV even for spallation secondaries such as boron. 
  
\section{Summary}
{ Noting previous work that suggests cosmic rays escape the disk relatively close to their sources (as compared to the radial scale of the disk), we have suggested that the grammage traverse by CR and the maximum energy to which they are accelerated both may vary with location in the disk.  Inverse correlation between grammage and maximum energy  would help explain the difference between secondary and primary spectra without invoking an energy-dependent escape rate.
The specific model analyzed here, in which ion-neutral damping sets the maximum energy,  suggests that the secondary spectrum may become more like the primary spectrum at higher energies, where ion neutral damping may play less of a role.}

 I thank Drs. Biman Nath,  E.N. Parker, and Noemie Globus for helpful conversations and comments on the manuscript. I acknowledge support from the ISF,  including an ISF-UGC grant, the Israel-US Binational Science Foundation, and the Joan and Robert Arnow Chair of Theoretical Astrophysics.
   
 \section{\bf References}






\noindent The AMS02 Collaboration, 
2016, PRL, 117, 1103
\noindent Bell, A.R.  1978, MNRAS 182, 443


\noindent   Cowsik, R., Madziwa-Nussinov, T., 2016, ApJ, 827, 119

 \noindent  Cowsik, R. and Wilson, L.W. (1973) Proc. ICRC, (Denver), 1, 500

\noindent D'Angelo, M., Blasi, P., \& Amato, E. (2016)  Phys. Rev. D, 94, 083003

\noindent Ellison, D.C. \& Eichler, D. 1985, PRL, 55,2735

 


\noindent  Drury, L. O'C, Duffy, P. and Kirk, J.,   1996, Astron. and Astroph., 392, 1002

\noindent Ellison, D.C., Drury, L. O 'C., Meyer, J.P.  1997,  ApJ, 487, 197

\noindent  Fiorito, R,. Eichler, D. and Ellison, D.C. 1990, ApJ, 364, 582

\noindent  Gupta, S., Nath, B. B., Sharma, P., Shchekinov, Y., 
2016 MNRAS, 462, 4532 

\noindent Kulsrud, R. and Cesarsky, C.J ., 1971, ApL 8, 189

\noindent Lagage, P. O.  and Cesarsky, C. J., 1983 Astron. and Astroph. 118, 223L

\noindent  Parizot, E. and Drury, L.,, 2000, Astron. and Astroph. , 356, L66

\noindent Parker, E.N.  1969, Space Science Reviews, 9, 65 
 http://adsabs.harvard.edu/abs/1969SSRv....9..651P

\noindent  Gupta, S., Nath, B. B., Sharma, P., Shchekinov, Y., 
2016 MNRAS, 462, 4532 









\end{document}